# An assessment of Sentinel-1 radar and Sentinel-2 multispectral data for remote archaeological investigation and preservation: Qubbet el-Hawa, Egypt.

O'Neill, C[1]., Bommas, M[2].

[1]Macquarie Planetary Research Centre, Department of Earth and Environmental Science, Macquarie University, Sydney, Australia.
[2]Macquarie University History Museum, Macquarie University, Sydney, Australia.

**Remote sensing for archaeological investigations using surface response is reasonably well established, however, remote subsurface exploration is limited by depth and penetration and ground resolution. Furthermore, the conservation of archaeological sites requires constant monitoring capability, which is often not feasible between annual field seasons, but may be provided by modern satellite coverage. Here we develop an approach using Sentinel-1 C-band radar backscatter, and Sentinel-2 multispectral data, to map and characterise the site of Qubbet el-Hawa, Egypt. The multispectral bands analysed show similar sensitivity to satellite imagery. However, the radar backscatter is sensitive to exposed known structures, as well as disturbances to soil textural/composition profile due to excavation/erosion. Sub-resolution features such as causeways manifest as a 'radar-break' in the backscatter - a discontinuity in otherwise continuous radar units. Furthermore, the finite subsurface response in the backscatter under the arid conditions of the site means we are able to delineate some shallow subsurface structures and map their orientation beneath the surface in areas not yet excavated. The sensitivity of Sentinel-1 backscatter to soil disturbance and human activity at Qubbet el-Hawa, and the short (~12 day) recurrence time of the satellites, makes it an important tool in heritage conservation.**

## Introduction

Geophysics and remote sensing are increasingly used in archaeological investigations for subsurface imaging of buried targets and structures, prior to excavation. These techniques save time, enabling broad subsurface characterisation and mapping, and restrict unnecessary exposure of a site to weathering and visitation.

Traditional field techniques, such as ground-penetrating radar (GPR), magnetic surveys, or resistivity, or Lidar topography surveys, require site access, transport of specialised equipment, specialised deployment over a field season. These requirements can restrict the use of these techniques at survey sites where logistical delivery of the equipment, safety concerns, or, more topically, travel restrictions, prevent deployment.

Satellite surveys face none of these field-site restrictions, and with numerous active missions, and recurrence times of satellite passes often less than 12 days, satellite mapping is increasingly used in the environmental and Earth science communities. The challenge facing satellite data are twofold: i) resolution is restricted by the average altitude of the satellite, and the precision of the sensor; and ii) penetration of the subsurface is limited.

The Sentinel-1 satellites orbit at an average of 693km altitude, and deploy a C-band synthetic aperture radar, with a central frequency of ~5.405GHz (Sentinel-1 technical guide). It is able to achieve resolutions of 5x5 m in strip-map mode, or, in its more common mode of interferometric wide-swath (IW), around 5x20m with a 250km wide swath.

Radar in archaeological settings is generally employed in on-site ground-penetrating radar surveys, and typically uses frequencies between 250-800MHz, depending on the depth of penetration required. Higher frequencies (1-1.5GHz) are often used in construction and engineering for shallow (< ~ 1m) imaging. The frequency used by Sentinel-1 is outside this range, but for targets such as dry sand, the subsurface transmission in not negligible, and some very shallow sensitivity to buried structure is possible (eg. Ghoneim et al., 2012). Primarily, though, surface structure will reflect radar depending on their orientation to the incoming pulse, and thus are amenable to imaging, and the interruption of strong radar reflectors (such as rock units) due to anthropological activity is mappable.

The Sentinel-2 satellites employ a multispectral imager to view the Earth in 13 bands in the visible, near and short-wave infrared in 290km swaths. The resolution is 10m for bands 2,3,4 and 8 (blue (492nm), green (559nm), red (665nm) and near-infrared (833nm), respectively), or 20m for the SWIR bands 11 (1610nm) and 12 (2189nm) employed here. The bands are sensitive to soil composition (eg. clay content, Bousbih et al., 2020), texture (Vadour et al., 2019), vegetation, and brightness/salinity, or moisture variations (Tagadoshi et al., 2019). Variations due to rock-soil contrast, soil disturbances, and soil texture may have archaeological significance.

Here we explore the capability of these techniques to characterise and map the archaeological site of Qubbet el-Hawa in Aswan, one of the best protected archaeological sites in Egypt, and ascertain its capability in monitoring soil disturbance and anthropogenic activity for heritage conservation. Qubbet el-Hawa is located opposite a police station and the local inspectorate, and therefore highly secured day and night. The characterisation and mapping of isolated locations would have to be assessed on their exposure to potential threats to the preservation of cultural heritage, and illicit digging. Importantly methods are needed for their remote monitoring, and here we additionally assess Sentinel-1 and 2 data for monitoring anthropogenic disturbance in a secure, well-characterised archaeological site.

# Site description

Qubbet el-Hawa (engl. 'The Dome of the Winds') is the name of an ancient Egyptian cemetery on the west bank of the River Nile, opposite Aswan. Often referred to as the Tombs of the Nobles, the rock-cut tombs at Qubbet el-Hawa were occupied by the administrative elite and priests that lived on the Island of Elephantine, north of the 1st Cataract. Despite the fact that the Island of Elephantine was inhabited around 3000 BCE, the earliest occupation of Qubbet el-Hawa as one of its burial places dates to the Old Kingdom (c. 2450 BCE). Tombs dating to the Middle Kingdom between 2200-1860 BCE and later are also present, in addition to continuous usurpation especially of the rock-cut tombs at the upper terrace. This makes Qubbet el-Hawa one of the richest and diverse cemeteries of Egypt, continuously occupied well into early Islamic periods. Here, governors of the area of Aswan who oversaw ancient Egypt's trade with Africa both along the Nile and via desert routes have first been buried at least since the reign of King Pepi II (c. 2278-2184 BCE). The region of Aswan is the least fertile in Egypt with granite quarries in the east and sandstone in the west reaching the Nile at Aswan. As a result, the ancient population relied upon international trade with Africa, customs and administration, and quarrying which generated income that was readily displayed in rock tomb architecture as well as wall decoration. However, until recently, lower administrative ranks had been missing from the archaeological record in the region of Aswan.

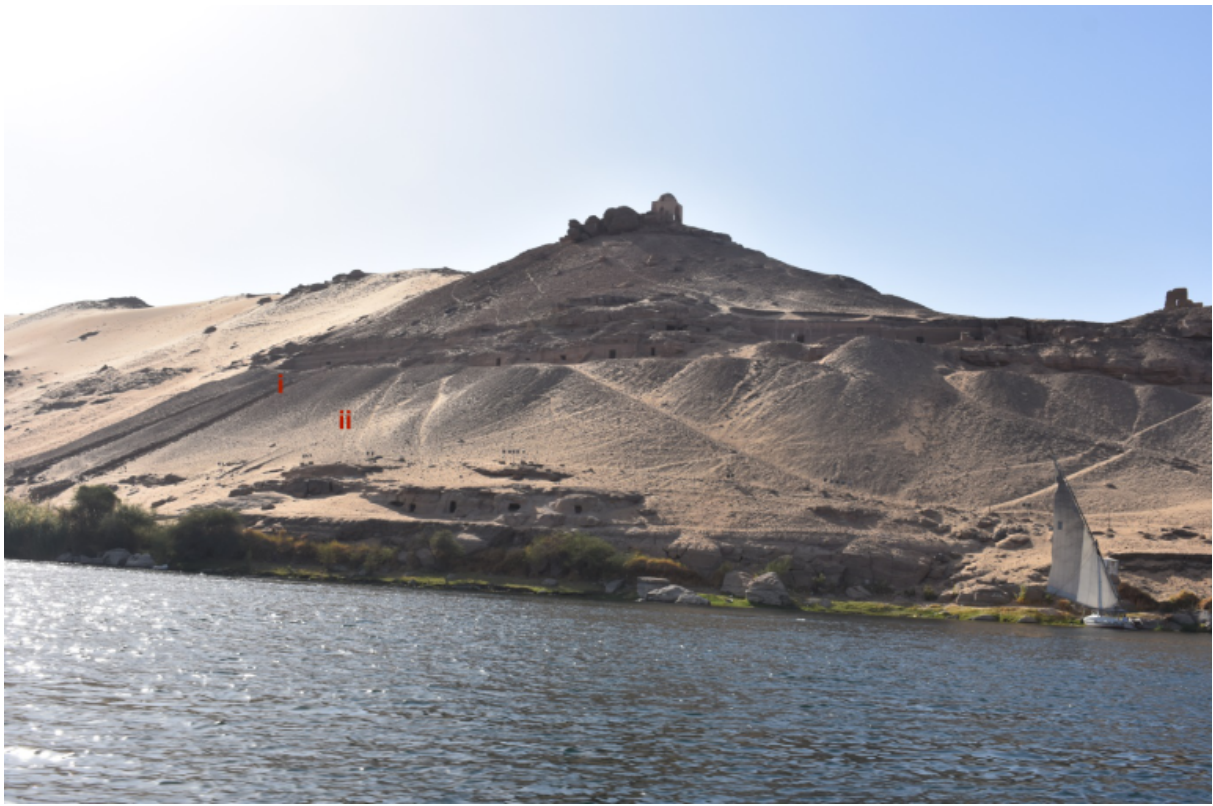

Fig. 1 Qubbet el-Hawa as seen from the east including the tomb of Sheikh Ali Abu el Hawa, transferred into a signalling station during the 19th Century, called 'the Dome'. To the left two monumental causeways lead to the rock-cut tombs of Sabni and Mekhu at the upper terrace (i), the causeway further to the right leads to the Old Kingdom tomb of Sobekhotep (ii), all belonging to governors of Elephantine.

Hidden underneath the sand, rock tombs of the administrative elite lined up between the tombs of Sabni and Mekhu (i) in the south (Figure 1) and the tomb of Sarenput I (Figure 3 (v)) in the north were first discovered in 1885. The Old Kingdom burials have been studied in full detail since the 1950s and were published in 2008 (Edel, 2008). The lower cemetery at Qubbet el-Hawa, on the other hand, was only discovered in 2016 for the joint University of Birmingham-Egypt Exploration Society Qubbet el-Hawa Research Project (QHRP), now a joint Macquarie University-Egypt Exploration Society expedition. Fieldwork at the site has so far concentrated on two areas: the discovery of nine mastaba tombs with bench-like mudbrick superstructures in the lower necropolis; and the discovery of three causeways leading to the upper terrace. It is here where burials of individuals of lower administrative ranks have finally come to light (Bommas, 2016). Most of the archaeological remains are hidden under massive layers of non-stratified rubble and wind-blown sand which prevented exploration of this area in the past. Apart from standard excavation methods and the application of intra-marginal context archaeology to assess the immediate environment of the mastaba tombs (Bommas, 2020), this detail-focussed approach does not allow for conclusions about patterns of occupancy of this otherwise virgin site. Also, Egyptological research and considerations based on comparisons with excavated contemporary sites, such as Beni Hassan in middle Egypt, do not reveal the full extent nor the overall dimensions of Qubbet el-Hawa beyond the border of current work permits and concessions received from the Government of the Arabic Republic of Egypt since 2015. Traditional survey methods to further assess Qubbet el-Hawa, especially with regards to the preservation of cultural heritage in an area which suffers from looting especially south of the tombs and Sabni and Mehu, are not practical due to the massive layers of aeolian sand which constantly re-shape the landscape since the past 2000 years after human activity in the area decreased.

Due to the landscape being covered by rubble (Figure 2), subsurface imaging such as through Ground Radar Penetration is impaired by the mixture of particle sizes underground which will confound GPR signal. Also, the Lower Necropolis built into the slope of Qubbet el-Hawa is often too steep to allow for operating laser scanners popular in Egypt such as the Leica ScanStation C10, due to the light and shifting wind-blown sand in the area.

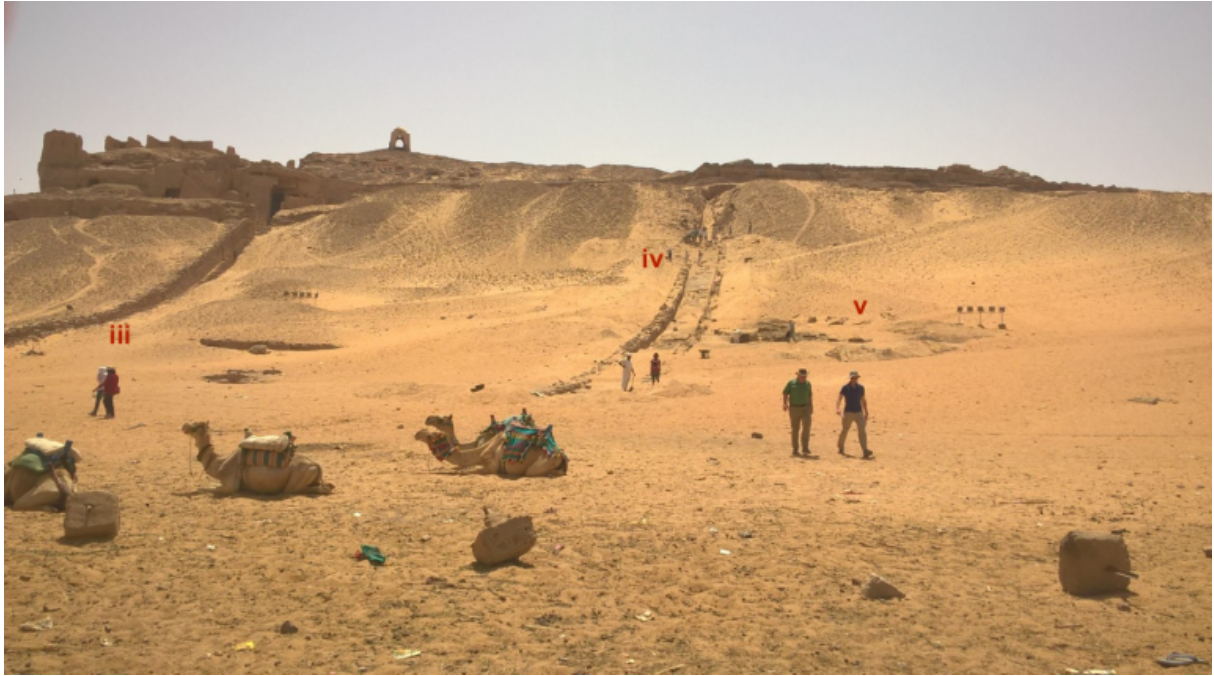

Fig.2 Overview of the current concession of QHRP. The causeway of Khunes is shown on the left (iii), in the centre the causeway 'Site A' leads to a rock-cut tomb below the upper terrace (iv), discovered in 2016. Darker areas denote rubble left from previous excavations of the upper terrace in the 1950s. The structures to the right of 'Site A' are remains of a mastaba necropolis, after cleaning in 2018 (v). View from the east. The diagonal line crossing 'Site A' is a modern camel track.

Remote sensing provides fresh opportunities to assess the scale and subsurface configuration of Qubbet el-Hawa (Figure 3). The Sentinel 1 and 2 constellations in particular have the potential to provide fresh insights into buried constructions, graves, and soil disturbance. In a dry sandy environment, Sentinel-1 C-Band radar has been shown to be sensitive to buried structures within the top ~5cm (Ghoneim et al., 2012) - of relevance to archaeological excations at Qubbet el-Hawa, as well as variations in soil texture, and moisture. Sentinel-2 provides a multispectral composites of the site in visible, near infrared (NIR) and short wave infrared (SWIR). The bands are variably sensitive to soil composition (eg. Van der Meer et al., 2014), soil-rock interfaces and porosity (Aliero et al., 2018). Qubbet el-Hawa is characterised by an arid, dry environment, with no vegetation cover at the site - but significant topographical and structural complexity, as well as strongly variable soil textures (gravel to sand). We assess the sensitivity of these sensors to delineating structures of archaeological significance, and anthropogenic influence, in this environmental setting in the following sections.

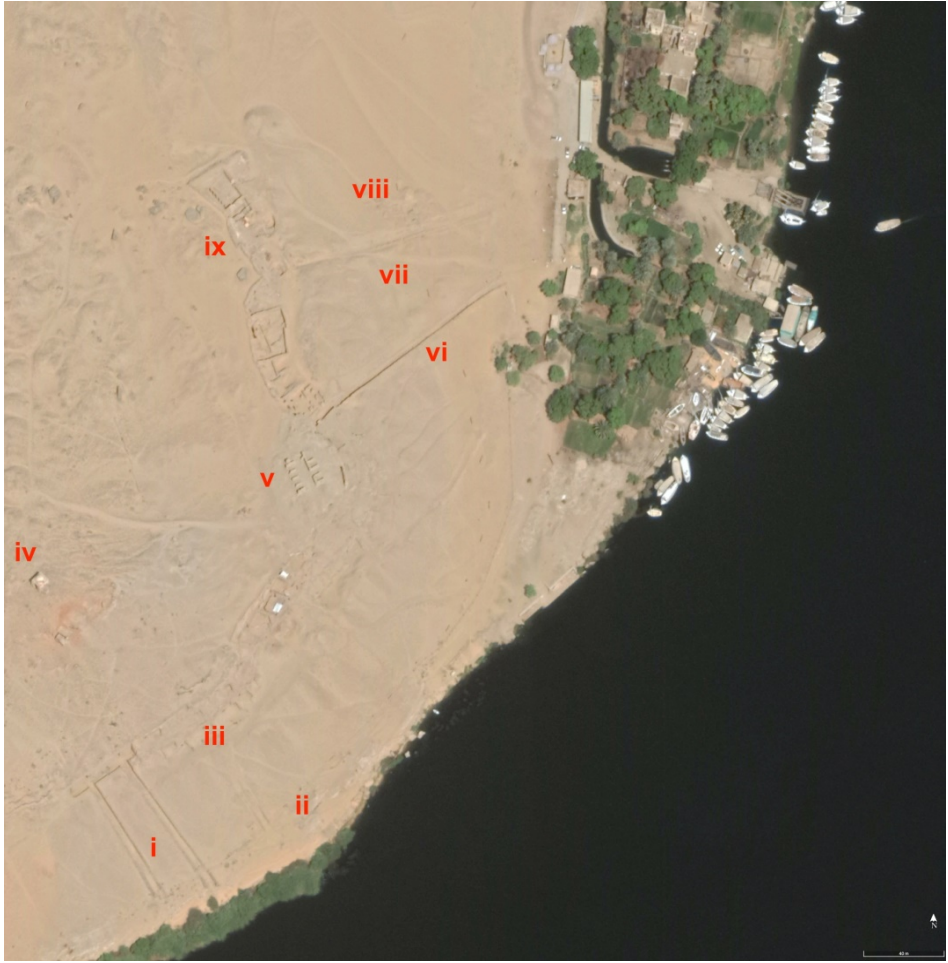

Figure 3. Satellite image of the Qubbet et Hawa archaeological site, Aswan (from Google Earth). Labelled features are i) Tombs of Sabni and Mekhu, and causeway, ii) Causeway of Sobekhotep, and iii) Tomb of Sobekhotep, iv) Hilltop, v) Tomb of Khunes, vi) Causeway of Khunes, vii) Site A excavation, and causeway, viii) Tomb of Sarenput I (causeway to the north partly excavated, but currently subsurface), and ix) Monastery of St Anthony.

## Methods

We extracted Sentinel-1 radar data for the periods 02/05/2017, 29/03/2020, and 10/04/2020, in dual polarisation (VV+VH), from the Copernicus hub (https://scihub.copernicus.eu/). We used ESA SNAP (http://step.esa.int) software and Sentinel-1 toolbox, and the gdal suite, to apply orbital corrections, radiometric calibration, terrain flattening, and terrain corrections (using SRTM DEM data). The data was imported into a python environment using gdal python bindings, and cropped and projected to WGS84 (EPSG:4326) using gdal_warp, and the radar bands extracted. The data was cropped to =/- 2 standard deviations to eliminate outliers, and regridded.

Sentinel-2 multispectral data were taken from a single swath of the area on 05/04/2020. Bands 2, 3, 4 and 8 were extracted from the 10m data subset, and bands 11 and 12 (SWIR) from the 20m data subset. In addition the total colour image (TCI) band was extracted to overlay the data on. This data was again regridded, and embedded into a kml file for viewing on google Earth.

## Results

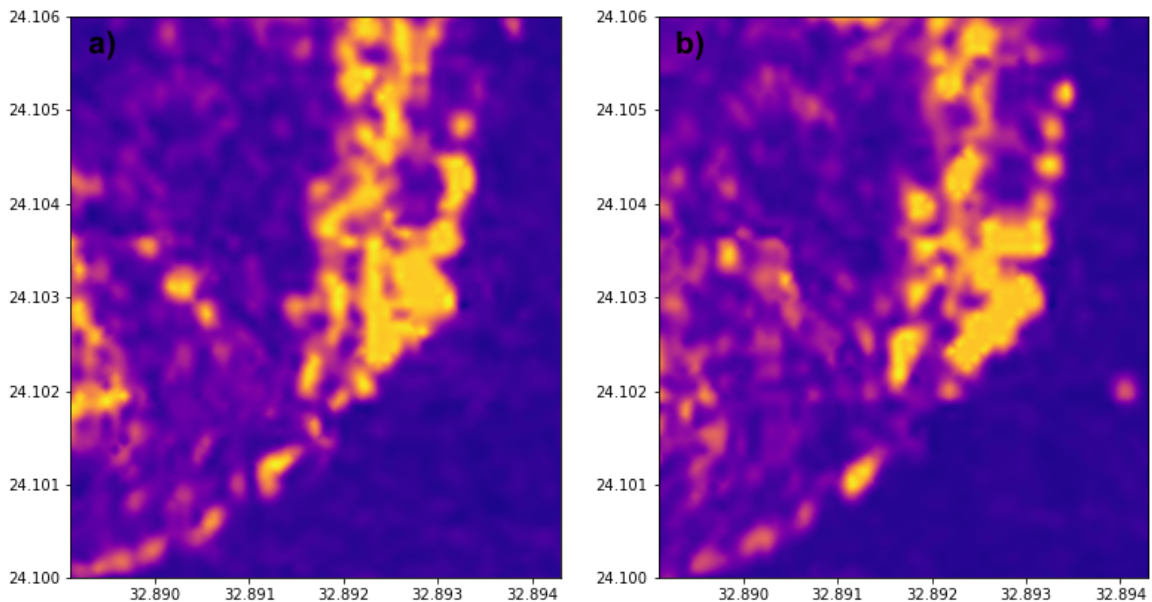

Figure 4. Sentinel-1 IW radar swaths from March, 2020. Left shows VH polarisation, right shows VV polarisation. Anthropogenic features (building roofs, boats) show up as strong reflectors in both images on the right, as does hard rock outcrop (left). Monastic ruins (centre) show as variable strength reflections depending on the polarity, and orientation to incident radiation.

Sentinel-1 IW backscatter is sensitive to strong radar reflectors (Figure 4) - including anthropogenic features such as roof tops and boats. The distribution of canals, fields, and trees in the urban area of Figure 3 moderates the backscatter intensity, though in different ways in each polarisation. Exposed rock ridges (left of Figure 4) display a strong backscatter intensity, although the orientation of the ridge exerts a strong control; the west of the ridge is high intensity in VV, the east shows a low backscatter shadow (Figure 4).

Known archaeological ruins, such the site of St Anthony's Coptic Monastery, and the Tombs of the Nobles, present as distinguishable features in the backscatter, though their coherency is broken up by changing radar incidence angles due to variable wall orientation.

Variations in backscatter at different times can arise from satellite position with respect to ground orientation of features, anthropogenic changes (construction, moving boats), or natural changes (vegetation differences, shifting sand surface). To assess the effect of these, and of the processing sequence we use, two sets of images are shown for 2017, and 2020. The latter has had the processing sequence described in the methods applied, the former represents raw data. The major backscatter features are seen in both datasets. The ruin outlines are more apparent in the processed 2020 data, although some smaller-scale archaeological features (eg. causeways) present more strongly in the raw data.

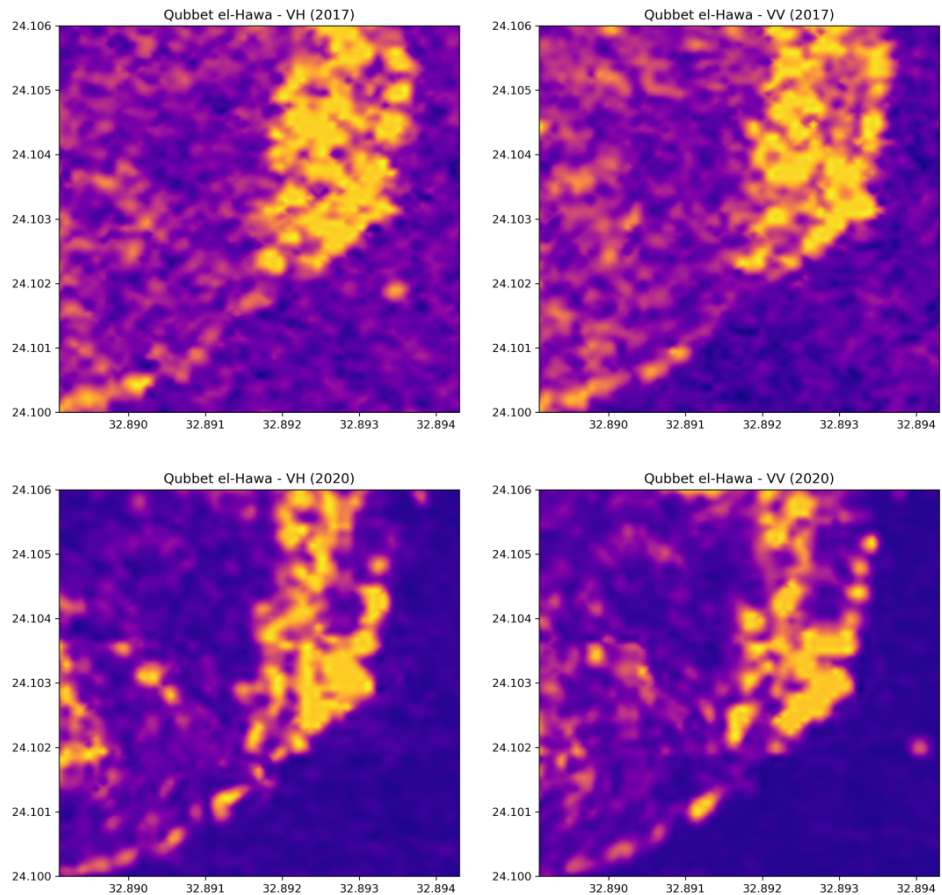

Figure 5. Comparison of raw Sentinel-1 backscatter in VH and VV polarisations for 2017, and corrected March 2020 images. The outline of the monastic ruins are observable in both sets of images.

The Sentinel-2 data for the bands with 10m resolution are plotted in Figure 6. The bands (B2-4, B8) are the BGR (blue-green-red) and 833nm near IR bands, respectively, and are sensitive to soil and surface variations, and soil-rock contrast.

Urbanised features show up strongly in these bands, and the contrast between building structures, playing fields, and canal networks is fairly apparent. The vegetation patterns to be strong absorbers (i.e. darker), and in individual trees and groves are identifiable. Water tends to be a low in RGB and NIR channels: canals and water bodies in the urbanised zone are particularly clear in the NIR. The exposed rock ridges are distinctive, as are the monastic ruins, and excavated archaeological features such as walls and ramparts.

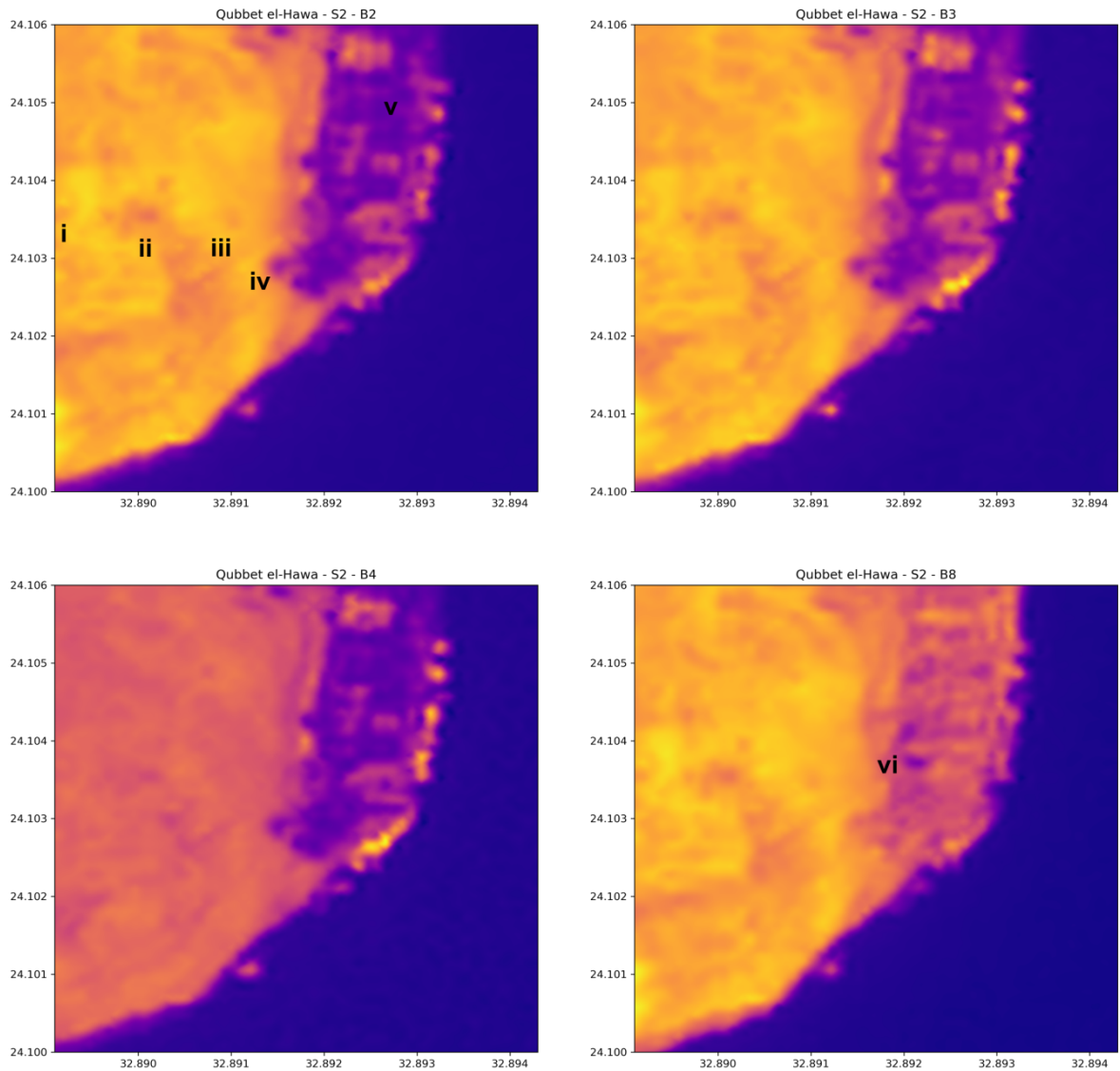

Figure 6. Sentinel-2 multispectral imager data for Qubbet el-Hawa. Shown are the 10m resolution bands 2 (blue, 492nm), 3 (green, 559nm), 4 (red, 665nm) and 8 (near-infrared, 833nm). i) Rocky hilltop, ii) the site of the Monastic ruins, iii) excavated Causeway of Khunes, iv) small grove of trees, v) urbanised strip; outline of building structures and grassed areas is apparent, vi) canals show are lows in the NIR.

Short-wavelength IR (bands 11-12) at 20m resolution show similar features, although with less resolution (Figure 5). Bands 11 and 12 are often used as a composite to map geology, and, together with B2 or B8, have been used in geology composite images (eg. Fal et al., 2018). Figure 7 shows the drop in resolution at this scale for the 20m datasets, although the major features identified in Figure 6 are still largely apparent. Small scale details like the causeway structure (see Figure 6) are still identifiable, but are heavily blurred, and the images would need significant enhancement and sharpening for use at archaeological-dig scales.

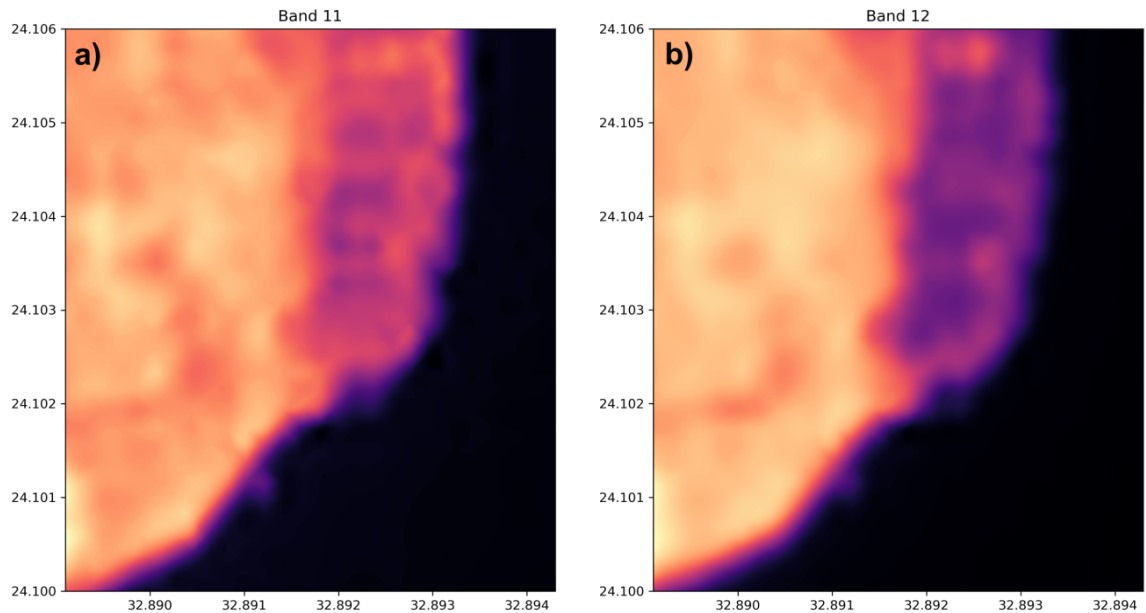

Figure 7. Sentinel-2 multispectral imager data for Qubbet el-Hawa, for SWIR bands 11 (1610nm) and 12 (2189nm), which have 20m resolution. Exposed archaeological structures have a strong response at these wavelengths, although with less definition than the 10m resolution bands.

# Discussion

Small-scale features at the level of an archaeological dig site are generally mapped using ground-techniques, such as GPR and other geophysical approaches, or detailed aerial/satellite imagery. Here we have shown that the Sentinel-1 and Sentinel-2 satellite constellations can be used to map and identify archaeological targets that themselves are below the nominal resolution of these datasets (~20m and ~10m respectively).

In the case of Sentinel-1 radar backscatter intensity, the one of the dominant mechanisms of backscatter for engineered structures is ground-wall double reflections (Koppel et al., 2017). This means the response of features such as walls is sensitive to inclination of the signal (and thus satellite position), the orientation of the wall, and also polarisation. In Figure 6 we have plotted the VV and VH polarisation backscatter intensities for the same swath (April, 2020). This swath was taken on an ascending orbit. Both polarisations detect different aspects of the structures observed; generally VH is sensitive to W-NW trending walls, VV to N-NE features. The strength of the signal is contingent on many factors - the texture of the ground, size of the structure and the contiguity, and local terrain variations.

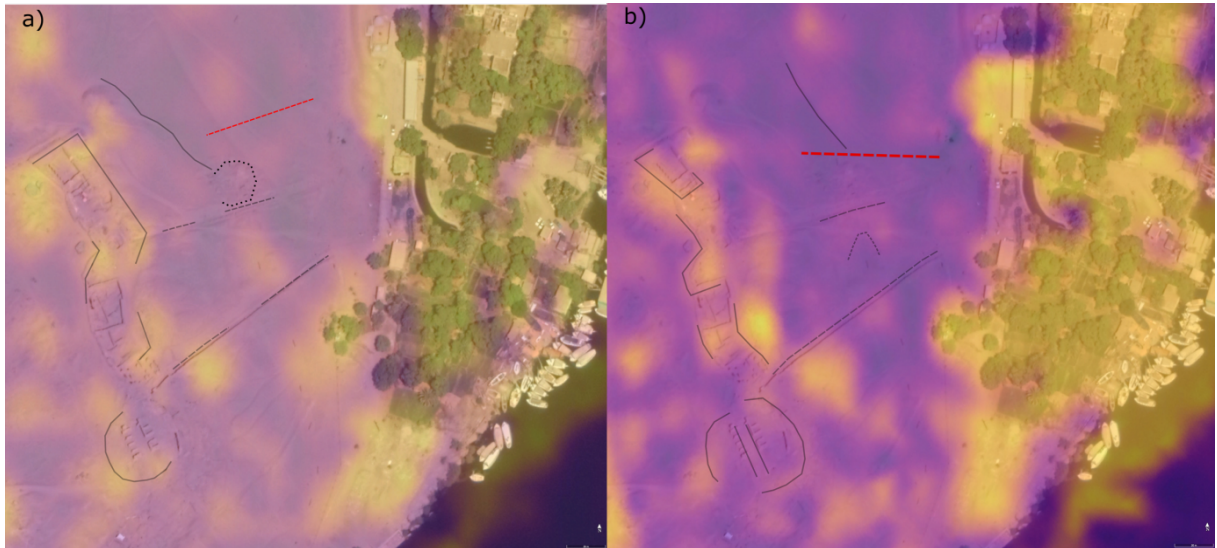

Figure 8. VH (left, a) and VV (b) polarisations draped over satellite imagery from GoogleEarth. Colorscale represents intensity of backscattered radar, and is sensitive to target orientation with respect to the satellite. Exposed archaeological structures with surface expression give strong responses, although the magnitude is dependent on the orientation. Examples are highlighted with dark solid lines. Some structures (walls, ramparts - dashed lines) are mappable due to their radar shadows breaking the coherency of strong backscatter targets (dashed lines). Red dashed lines indicate radar breaks without surface expression; in a) the break follows the buried Causeway of Sarenput I, in b) the radar break signal is not known. Other variations are due to soil composition or texture variations (eg. camel track highlighted in the north of the image).

In many cases, however, the presence of sub-resolution features is not directly detectable in the return backscatter intensity, but rather as a break in the continuity of other features, such as local soil variations. In the case of two excavated causeways of Qubbet el-Hawa, these structures are characterised by a break in the intensity of VV or VH backscatter of soil units, and often a change in the dominant backscattered polarisation, from VV on the south side of these structures to VH in the north, possibly due to reduced ground-wall reflections. Interestingly, a radar break appears to be associated with the currently buried Causeway of Sarenput I. This causeway to the north is indicated by a red dashed line in Figure 8a, but is not currently exposed. It has been excavated partially in its eastern extent, but the radar break delineates its inferred orientation.

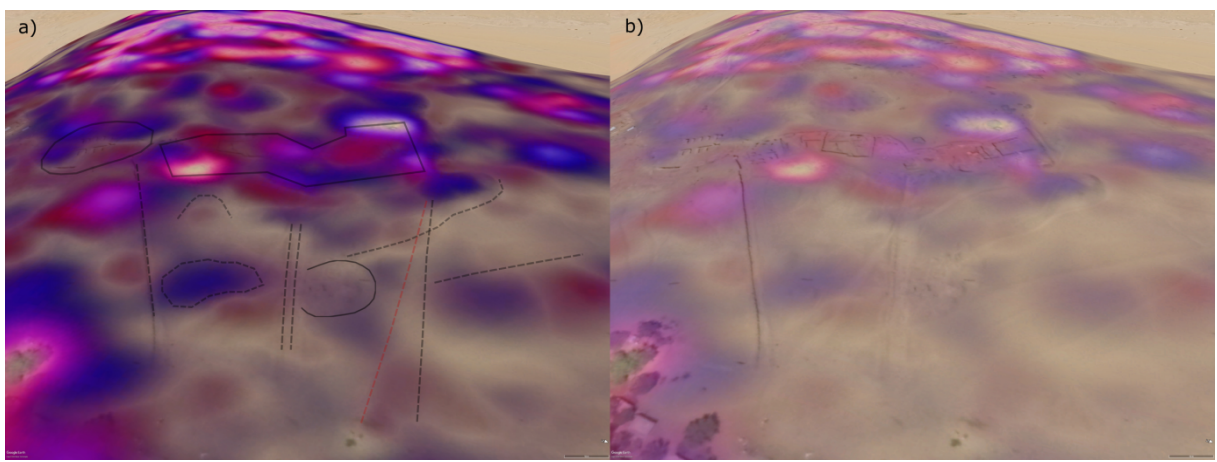

Figure 9. Radar backscatter from Sentinel-1 in April, 2020. Red channel represents VV polarisation, blue channel VH (green channel is set to null, and opacity is a sine function of intensity). The image has been draped over terrane on GoogleEarth, and a variable transparency applied to see the satellite imagery underneath (b). Black lines in (a) delineate major radar features. Solid lines are major archaeological features delineated by intensity variations, dashed lines represent features delineated by radar shadows, or unknown variations (blue region in lower centre). These include major ramparts (observed in (b)), camel tracks (dashed lines on the right of (a)), and a small alluvial fan (red region encompassed by a small dashed parabola) with textural variation to surrounding sand. The red dashed lines indicate a crypto-radar break seen in Figure 8b that has no known surface expression.

Figure 9 shows an RGB representation of the polarised Sentinel-1 radar data, with VV shown as the red channel, and VH as the blue (green is null, and a sine function is used to map opacity to aid visualisation). The data is again from April 2020, for an ascending orbit. The major archaeological features (Monastery of St Anthony, Tombs of the Nobles) are shown as variable polarisation responses around their boundaries, which are reasonably well-delineated. However, major excavated features of this commission, such as the causeways, are largely identified by radar shadows, breaks in the soil polarisation, and inversions in the dominance of VV-VH across the structure. Variations in the soil texture plays a role in the intensity variation; a small alluvial fan forming on the steep slope beneath the monastery is evidence as a small VV high (parabola in Figure 7), and a high VH feature (encircled blue feature) is associated with disturbed soil.

Our results indicate a limited - but finite - response of subsurface structures in the Sentinel-1 data - probably due to contrasts within the upper 5cm. The ground at the site is sandy and dry - which is ideal for radar penetration - but buried structures much deeper than ~5cm are not expected to be delineated. This suggests a number of complementary strategies to assist in subsurface detection, including: i) a combination of L-band data which can achieve greater penetration in dry sand (eg. Ghoneim et al., 2012), ii) coherency mapping over time to identify fixed subsurface features under mobile, wind-driven sediments, iii) tracking moisture indices over time, which may be sensitive to soil pore water flow, and buried structures, or iv) using dInSAR to delineate surface deformation over time, which again may show sensitivity to subsurface structure.

Lastly, the variation in polarisation response is strongly affected by changes in surface soil texture, particularly associated with excavation. Given the short recurrence time of Sentinel-1 data acquisition (< 12 days), this suggests that radar backscatter monitoring could have use in archaeological heritage monitoring, tracking excavation scope, and illegal excavation activity. In Qubbet el-Hawa, textural and compositional variations in the soil lead to distinct backscatter anomalies for perturbed soil, including at older dig sites (eg. the excavated mastaba necropolis, and near the Causeway of Khunes, see Figure 2). Sediment transport downslope in an alluvial fan is also discernible in Figure 9. This lends support to the use of Sentinel-1 radar backscatter in monitoring soil disturbance at remote archaeological sites, for heritage conservation.

# Conclusions

The use of satellite mapping for archaeological surveying is attractive for its temporal coverage and accessibility, compared to traditional ground techniques such as geophysical surveys. However, they face a challenge in spatial resolution, and also ground penetration. Here we have assessed utility of Sentinel-1 radar data, and Sentinel-2 multispectral imagery for mapping the archaeological site of Qubbet el-Hawa in Aswan, Egypt. The radar and multispectral data are able to delineate excavated archaeological structures, and also soil textural variations associated with excavation, or natural processes (eg. alluvial fan formation). Breaks in the coherency of the Sentinel-1 backscatter signal allow the delineation of some structures beneath very shallow sand, that correlate in orientation with past excavations - including the Causeway of Sarenput I in the north - although a nuanced approach to the contributions to radar backscatter variations and their interpretation is clearly necessary. An understanding subsurface signal contributions to C-band radar requires fairly detailed knowledge of surface soil texture and composition, but the sensitivity of Sentinel-1 backscatter to these factors demonstrates its importance as a tool for remote monitoring of soil disturbance at archaeological sites for heritage conservation.